\documentclass[abstract,a4paper,fleqn,10pt]{scrartcl}
\pdfoutput=1
\usepackage[left=1.25in, right=1.25in, bottom=1.25in,
top=1.25in]{geometry}

\usepackage{hyperref}
\usepackage{graphicx}
\usepackage{siunitx}
\usepackage{commath}
\usepackage{color}
\usepackage{amsmath}
\usepackage{pifont}

\usepackage{
  pgf,
  tikz}
\usetikzlibrary{
  arrows,
  decorations.markings,
  positioning,
  calc
}

\hypersetup{
  pdfauthor={Stephen Jones, Silvan Kuttimalai}, 
  pdftitle={}
}

\usepackage{authblk}

\author[0]{Stephen Jones}
\affil[0]{Max Planck Institute for Physics, F{\"o}hringer Ring 6,
  80805 M{\"u}nchen, Germany}

\author[1]{Silvan Kuttimalai}
\affil[1]{SLAC National Accelerator Laboratory, Menlo Park, CA 94025, USA}

\title{Parton Shower and NLO-Matching Uncertainties in Higgs Boson Pair
  Production}

\titlehead{MPP-2017-249\\SLAC-PUB-17190}

\begin{document}

\maketitle

\begin{abstract}
  We perform a detailed study of NLO parton shower matching
  uncertainties in Higgs boson pair production through gluon fusion at the
  LHC based on a generic and process independent implementation of NLO
  subtraction and parton shower matching schemes for loop-induced
  processes in the Sherpa event generator. We take into account the
  full top-quark mass dependence in the two-loop virtual corrections and
  compare the results to an effective theory approximation. In the
  full calculation, our findings suggest large parton shower matching
  uncertainties that are absent in the effective theory approximation.
  We observe large uncertainties even in regions of phase space where
  fixed-order calculations are theoretically well motivated and parton
  shower effects expected to be small. We compare our results to NLO
  matched parton shower simulations and analytic resummation results
  that are available in the literature.
\end{abstract}

\section*{Introduction}


In the Standard Model (SM) the production of a pair of Higgs bosons
at a hadron collider proceeds dominantly through the annihilation of gluons. 
As there is no direct coupling between the Higgs boson and gluons this process
is mediated by intermediate massive quark loops. The top-quark loop contributions 
dominate by far, due to the large Yukawa coupling.
Bottom-quark loops only contribute at the \SI{1}{\percent} level to
the total cross section at leading order (LO) and can thus safely be
neglected in most situations. The scattering amplitudes relevant for
the calculation of the top-quark contributions up to next-to-next-to leading
order (NNLO) are known in the approximation of an infinitely
heavy top-quark (commonly referred to as the HEFT approximation)
\cite{Dawson:1998py,deFlorian:2013uza,deFlorian:2013jea,Grigo:2014jma,Grigo:2015dia,deFlorian:2016uhr}.
However, the validity of this approximation is questionable for 
Higgs boson pair production due to the large momentum transfer required to produce the Higgs bosons.
Techniques to systematically improve upon it were extensively studied in
\cite{Frederix:2014hta,Maltoni:2014eza,Grigo:2013rya,Grigo:2014jma,Grigo:2015dia,Maierhofer:2013sha}.
The full result, which is exact in the mass of the top-quark,
was known only to leading order (LO) \cite{Eboli:1987dy,Glover:1987nx,Plehn:1996wb} 
until recently. This is due to the complexity of computing
the next-to-leading order (NLO) virtual corrections which feature
two-loop, four-point integrals with both massive internal propagators and massive
external lines. They have recently been calculated through the
numerical evaluation of all relevant two-loop integrals as part of the
full NLO calculation in \cite{Borowka:2016ehy,Borowka:2016ypz}.

At small transverse momenta $p_\perp^{HH}$ of the
Higgs boson pair, the accuracy of any fixed-order calculation is spoiled by
the presence of large logarithms of the form
$\alpha_s^{n} \bigl[\log(p_\perp^{HH}/m_{HH})\bigr]^{m}$. They can
be resummed to all orders using analytical resummation techniques 
which have been applied to Higgs boson pair production in
\cite{Ferrera:2016prr}. Alternatively, parton shower simulations can
be employed. In addition to providing a reliable transverse momentum
spectrum at small $p_\perp^{HH}$, they also provide results that are
fully differential in the kinematics of any soft and collinear QCD
radiation. Standard techniques exist for the consistent matching of
NLO fixed order calculations to parton shower simulations
\cite{Frixione:2002ik,Nason:2004rx}. They were recently applied to
Higgs boson pair production in reference \cite{Heinrich:2017kxx}, where the
MC@NLO and POWHEG matching techniques were used to combine the
fixed-order NLO calculation with the Pythia parton shower
\cite{Sjostrand:2007gs,Sjostrand:2014zea}. The results of
\cite{Heinrich:2017kxx} suggest that the parton shower matching can
have sizeable effects not only in the region of small $p_\perp^{HH}$,
but also in the region of large $p_\perp^{HH}$, where one would expect
the fixed-order calculation to be reliable and the approximations
inherent to parton shower simulations to break down. These effects
even exceed the scale uncertainties of the fixed-order calculation.

In this publication we aim to critically assess the origin and size of
the aforementioned effects and associated uncertainties. For this
purpose we implemented a fully generic and process independent NLO
subtraction along with the corresponding parton shower matching techniques
for loop-induced processes in the Monte Carlo event generator Sherpa
\cite{Gleisberg:2008ta}. This allows us to perform our studies using
the two different showers that are implemented in Sherpa
\cite{Schumann:2007mg,Hoche:2015sya} within the same parton shower
matching framework.

This publication is structured as follows. In Section~\ref{sec:setup}
we describe in detail the setup of our calculation along with a brief
review of the MC@NLO matching technique and the parton showers we used
for our studies. We present the results of our simulations in Section~\ref{sec:results}, 
focusing on the origin and size of uncertainties
that are inherent to the matching technique applied. We also point out
crucial differences that arise when going from the HEFT approximation
to the full calculation. Our conclusions are presented in Section~\ref{sec:conclusions}.


\section{Calculational Setup}
\label{sec:setup}

\subsection{Fixed-Order NLO Calculation}

For the virtual two-loop amplitude we utilize the result of reference
\cite{Borowka:2016ehy,Borowka:2016ypz}, retaining the full finite
top-quark mass effects. This amplitude was obtained by numerically
evaluating all relevant 2-loop 4-point Feynman diagrams with up to 4
scales. We adopt the input parameters of reference
\cite{Borowka:2016ypz}, with
$G_F= 1.1663787 \times 10^{-5}\ \mathrm{GeV}^{-2}$, the mass of the
top-quark set to $m_t = \SI{173}{\giga\electronvolt}$, the Higgs boson
mass set to $m_H = \SI{125}{\giga\electronvolt}$, and their widths
neglected. We also adopt the choice of reference
\cite{Borowka:2016ypz} for the factorization and renormalization
scales $\mu_F=\mu_R=m_{HH}/2$. Perturbative uncertainties in the
fixed-order part of the calculation are estimated by independently
varying these scales through factors of \num{0.5} and \num{2}.
All studies are performed with hadronic center-of-mass energy $\sqrt{s} = \SI{14}{\tera\electronvolt}$.
The NLO virtual amplitude is provided in the literature in the form of
an interpolation grid in two Mandelstam variables, based on a fixed
number of pre-computed phase-space points \cite{Heinrich:2017kxx}. 
We extract the finite part of the UV renormalized virtual amplitude in the
Conventional Dimensional Regularization (CDR) scheme 
with residual IR singularities subtracted according to the Catani and Seymour 
scheme \cite{Catani:1996vz}, as required by the Sherpa event generator, 
using relations (2.5) and (2.6) of reference \cite{Heinrich:2017kxx}.


The leading order one-loop squared amplitudes for the Born process and
real emission contributions are provided by OpenLoops
\cite{Cascioli:2011va}. For the evaluation of tensor and scalar
one-loop integrals, we employ the Collier library
\cite{Denner:2016kdg}, CutTools \cite{Ossola:2007ax}, and OneLOop
\cite{vanHameren:2009dr,vanHameren:2010cp}.

For the regularization and numerical cancellation of infrared
divergences in the real-emission part of the calculation we employ the
dipole subtraction scheme of Catani and Seymour \cite{Catani:1996vz}.
We have re-implemented this scheme within Sherpa in a fully generic
and process-independent way for loop-induced processes. This
implementation is qualitatively equivalent to the implementation in
one of Sherpa's internal matrix element generators Amegic++
\cite{Krauss:2001iv,Gleisberg:2007md}, apart from the fact that color-
and spin-correlated amplitudes are to be provided externally through
generic interfaces. Through a dedicated interface to OpenLoops and the
aforementioned tools, NLO calculations for loop-induced SM processes
are thus fully automated (given the availability of the virtual
two-loop corrections) in Sherpa and will become available in a public
version of the code.

We have validated our implementation in Sherpa by comparing our results for the
total cross section, for the differential Higgs boson pair invariant mass
distributions, and for the differential single Higgs boson transverse
momentum distributions to those published in reference \cite{Borowka:2016ypz}.

\subsection{Parton Showers}
\label{sec:ps}

We consider two parton showers for matching to the fixed-order NLO
calculation. Both algorithms are dipole-type showers in which QCD
radiation is generated coherently off color dipoles spanned by pairs
of pre-existing partons. Both showers are publicly available as part
of the Sherpa event generator package.

Due to the dipole character of the parton showers, their splitting
kernels can be used for the purpose of fixed-order NLO subtraction,
thus simplifying the implementation of parton shower matching. The CS
shower \cite{Schumann:2007mg} directly uses the splitting kernels of
the original Catani-Seymour subtraction scheme for parton evolution.
The Dire shower \cite{Hoche:2015sya} uses splitting kernels that are
modified in such a way as to reproduce the collinear anomalous
dimensions of the DGLAP equations. For NLO matching to the Dire
shower, we use a modified version of the original Catani-Seymour
subtraction scheme that reflects these changes in the splitting
kernels \cite{Hoeche:DireSubtraction}. The kernels of both showers
approximate real emission amplitudes arbitrarily well in the limit of
soft and collinear momenta. Away from the soft and collinear regions,
however, they differ.

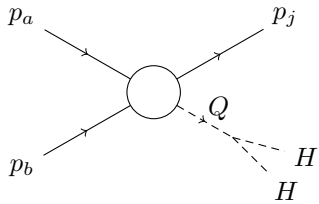
\begin{figure}
  \centering
  \tikzset{
  blob/.style=
  {
    minimum size=20pt,
    circle,
    draw=black, 
    fill=white,
    inner sep=0pt, 
    outer sep=0pt
  },
  vertex/.style=
  {
    minimum size=00pt,
    circle,
    draw=black, 
    fill=white,
    inner sep=0pt, 
    outer sep=0pt
  },
  gluon/.style=
  {
    decorate,
    draw=black,
    decoration={coil,amplitude=2pt,segment length=2.5pt}
  },
  real_scalar/.style=
  {
    densely dashed,
    thin,
    draw=black,
  },
  real_scalar_arr/.style=
  {
    densely dashed,
    thin,
    draw=black,
    postaction={decorate},
    decoration=
    {
      markings,
      mark=at position 0.5 with {\arrow[draw=black]{>}}
    }
  },
  flow/.style=
  {
    thin,
    draw=black,
    postaction={decorate},
    decoration=
    {
      markings,
      mark=at position 0.5 with {\arrow[draw=black]{>}}
    }
  }
}
  \begin{tikzpicture}
 \node [blob] (M)  at (1.5,0)  {};
 \node [on grid] (g1) at ($(M)+(-30:-2)$)   {$p_a$};
 \node [on grid] (g2) at ($(M)+(+30:-2)$)   {$p_b$};
 \node [on grid] (g3)  at ($(M)+(+30:2.0)$)   {$p_j$};
 \node [vertex]  (x)   at ($(M)+(-30:1.2)$) {};
 \node [on grid] (h0)  at ($(x)+(-15:1)$) {$H$};
 \node [on grid] (h1)  at ($(x)+(-45:1)$) {$H$};
 \draw [flow] (g1) -- (M);
 \draw [flow] (g2) -- (M);
 \draw [flow] (M)  -- (g3);
 \draw [real_scalar_arr] (M) -- (x) node[above, near end] (q2) {$Q$};
 \draw [real_scalar] (x) -- (h0);
 \draw [real_scalar] (x) -- (h1);
\end{tikzpicture}
  \caption{Definition of kinematic variables in the first emission.}
  \label{fig:ps-kinematics}
\end{figure}

A further crucial difference between the two algorithms is the choice
of evolution variable, which we generically denote by $t$ in the
following. The choice of evolution variable together with the shower
starting scale $\mu_\text{PS}$ dictates how much of the phase space
away from the soft and collinear regions is available to the parton
shower since the starting scale implements the following phase space
constraint:
\begin{align}
t < \mu^2_\text{PS}. \label{eq:mups-def}
\end{align}
For the discussion of the evolution variables we focus on the
first (hardest) emission in the production of a color-neutral final
state of invariant mass $Q^2=m^2_{HH}$. 
We illustrate the kinematics of the first emission,
producing a final state parton with momentum $p_j$ from the collision
of two incoming massless partons with momenta $p_a$ and $p_b$, in Figure~\ref{fig:ps-kinematics}. 
It is useful to consider the variables $v$
and $w$, which are closely related to the standard Mandelstam variables
$\hat{t}=(p_a-p_j)^2$, $\hat{u}=(p_b-p_j)^2$, and
$\hat{s}=(p_a+p_b)^2$:
\begin{align}
  v &= \frac{p_a p_j}{p_a p_b}=\frac{-\hat t}{\hat s}\geq 0, &
  w &= \frac{p_b p_j}{p_a p_b}=\frac{-\hat u}{\hat s}\geq 0.
\end{align}
Due to momentum conservation $\hat{s}+\hat{t}+\hat{u}=Q^2$ which implies,
\begin{align}
v+w &= (1-\frac{Q^2}{\hat s})<1 &\Rightarrow& & v w &< \frac{1}{4}. \label{eq:max-t-dire}
\end{align}

In terms of $v$ and $w$, the evolution variable in Dire is given by
\begin{align}
  \frac{t^\text{Dire}}{Q^2} = \frac{(p_a p_j) (p_b p_j)}{(p_a p_b)^2} = v w. \label{eq:t-dire}
\end{align}
The functional form \eqref{eq:t-dire} implies that $t^\text{Dire}<\frac{Q^2}{4}$
due to equation \eqref{eq:max-t-dire}. It follows that for a parton
shower starting scale of $\mu^2_\text{PS}=\frac{Q^2}{4}$, Dire behaves
like a ``power shower'' in the sense that it populates the full phase
space since $t^\text{Dire}<\frac{Q^2}{4}$ and thus \eqref{eq:mups-def} is
trivially fulfilled.

In the CS shower, the evolution variable is given by
\begin{align}
  \frac{t^\text{CSS}}{Q^2} = \frac{v w}{1-(v+w)}. \label{eq:t-css}
\end{align}
This implies that, for a given kinematic configuration,
$t^\text{CSS}$ is typically larger than $t^\text{Dire}$, such that for a given
value of $\mu_\text{PS}$ the emission phase space of the CS shower is more restricted than
that of Dire. It is worth noting that
$\mu^2_\text{PS}=\frac{Q^2}{4}$ in particular does not correspond to a
``power shower'' when employing the CS shower. This choice in fact
severely constrains the emission phase space, since $v+w$ can get
close to \num{1} and thereby give rise to large values of
$t^\text{CSS}$.

\subsection{NLO Parton Shower Matching}
\label{sec:nlops}

In the following we will focus on the MC@NLO matching prescription
\cite{Frixione:2002ik} using the notation of \cite{Hoeche:2011fd} with
no distinction between fixed-order NLO subtraction terms $D^{(S)}$ and
parton shower matching terms $D^{(A)}$ since we use the parton shower
splitting kernels both for parton evolution and for infrared
subtraction and keep all phase space constraints explicit. 
We thus denote the sum of subtraction terms as a function of
the real emission phase space by $D(\phi_R)$, where the real emission
phase space $\phi_R=\phi_B\times\phi_1$ can be decomposed into the
Born phase space $\phi_B$ and an extra one-particle emission phase
space $\phi_1$. We then define the fixed-order differential seed cross
sections $\bar B(\phi_B)$ and $H(\phi_R)$ in terms of the leading
order (Born) term $B(\phi_B)$, the UV-subtracted virtual corrections
$V(\phi_B)$, and the real-emission contributions $R(\phi_R)$ by
\begin{align}
  \bar B(\phi_B) &= B(\phi_B)+V(\phi_B)+\int D(\phi_R) \Theta(\mu^2_\text{PS}-t(\phi_R))\dif\phi_1\nonumber\\
                 &= B(\phi_B)+V(\phi_B)+I(\phi_B), \\
                   H(\phi_R) &= R(\phi_R) -
                               D(\phi_R)\Theta(\mu^2_\text{PS}-t(\phi_R))\ ,
                               \label{eq:mcnlo-seed-xs}
\end{align}
where $t(\phi_R)$ is the map from a kinematic real emission
configuration to the parton shower evolution variable $t$. The
Heaviside function $\Theta(\mu^2_\text{PS}-t(\phi_R))$ in
\eqref{eq:mcnlo-seed-xs} implements the constraint
\eqref{eq:mups-def}. For notational convenience, we will omit the
explicit $\phi_R$-dependence and write $t(\phi_R)=t$ in the following.
In terms of the quantities introduced above, the fixed-order total NLO
cross section is given by
\begin{align}
  \sigma_\text{NLO} = \int\bar B(\phi_B)\dif\phi_B + \int H(\phi_R)\dif\phi_R.
\end{align}
In MC@NLO, we generate events according to 
\begin{align}
    &\sigma_\text{MC@NLO} = \nonumber\\
    &\int \bar B(\phi_B)
      \underbrace{\left[\Delta(t_0,\mu^2_\text{PS}) + \int
      \Delta(t,\mu^2_\text{PS})\frac{D(\phi_B,
      \phi_1)}{B(\phi_B)}\Theta(\mu_\text{PS}^2-t)\Theta(t-t_0)\dif\phi_1\right]}_{\text{S-events}}\dif\phi_B\nonumber\\
    +&\int\underbrace{H(\phi_R)}_{\text{H-events}}\dif\phi_R,
  \label{eq:mcnlo}
\end{align}
where $t_0$ is the infrared cutoff scale of the parton shower and
the modified Sudakov form factor $\Delta(t_0,t_1)$, which gives the probability
for no emission to occur between scales $t_0$ and $t_1$ for the first parton shower
step, is given by
\begin{align}
  \Delta(t_0,t_1) &= \exp\left[ -\int^{t_1}_{t_0}\frac{D(\phi_R)}{B(\phi_B)}\dif\phi_1 \right]\\
  &= \exp\left[ -\int\frac{D(\phi_R)}{B(\phi_B)}\Theta(t_1-t)\Theta(t-t_0)\dif\phi_1 \right].
\end{align}
The first line of \eqref{eq:mcnlo} corresponds to events that have
Born kinematics at the level of the fixed-order seed event with weight
$\bar B$ (S-events). They either don't undergo any emission above the
infrared parton shower cutoff scale $t_0$ (first term in the square
bracket) or they undergo their hardest emission at some scale $t$
between $\mu^2_\text{PS}$ and $t_0$ (second term in the square
bracket). The second line of \eqref{eq:mcnlo} corresponds to events
with real-emission kinematics at the level of the fixed-order seed
event and weight $H$ (H-events). All events are treated further
by the parton shower precisely as in the leading order case,
apart from the S-events that haven't undergone any emission, which are
kept as they are.

Since the square bracket in \eqref{eq:mcnlo} integrates to \num{1},
the total cross section and any observable that is insensitive to QCD
radiation is unaltered in MC@NLO compared to the fixed-order NLO
result. In fact, it can be shown that a MC@NLO event sample will
reproduce the fixed-order NLO result event to order $\alpha_S$
relative to the Born for \emph{any} infrared safe observable
\cite{Frixione:2002ik}. The parametric NLO accuracy is therefore not
spoiled by the parton shower matching.

\subsection{Parton Shower Matching Uncertainties}
\label{sec:nlops_unc}

As stated in the previous section, NLO parton shower matching
according to the MC@NLO method preserves the parametric accuracy of
the fixed-order NLO calculation. Deviations from fixed-order results
can numerically be nonetheless significant \cite{Hoeche:2011fd}. Such
differences reflect genuine parton shower matching
uncertainties, they can be particularly prominent for observables that
are sensitive to real emission configurations and thereby
to the interplay between parton shower emissions and fixed-order real
emission configurations. We will therefore focus on the $p_\perp^{HH}$
distribution in the following section, comparing MC@NLO matched parton shower
simulations to fixed-order results with both the Dire and the CS
shower. 

In order to formally compare the MC@NLO result to a
fixed-order prediction for this spectrum, we first consider a generic
observable $\mathcal{O}$ that is insensitive to kinematic Born
configurations. For such an observable we need to take into account
H-events and parton shower emissions off S-events. At order $\alpha_S$
relative to the Born we have
\begin{align}
  \langle\mathcal{O} \rangle &= \int\bar B(\phi_B)
                               \Delta(t,\mu^2_\text{PS})\frac{D(\phi_B,
                               \phi_1)}{B(\phi_B)}\Theta(\mu_\text{PS}^2-t)\mathcal{O}(\phi_R)\dif\phi_B\dif\phi_1\nonumber\\
  &+\int H(\phi_R)\mathcal{O}(\phi_R)\dif\phi_R, \label{eq:real-obs}
\end{align}
where the first integral corresponds to S-events in which the parton
shower has generated a non-vanishing value of $\mathcal{O}$ and the
second integral corresponds to H-events, where a non-vanishing value
of $\mathcal{O}$ is implied by the real-emission kinematics of the
fixed-order seed event. In the tail of the distribution where we can
neglect the Sudakov suppression and set $\Delta=1$, we obtain after
plugging in the definition of $H$:
\begin{align}
  \langle\mathcal{O} \rangle &=\int
                               \left[\bar B(\phi_B)-B(\phi_B)\right]\frac{D(\phi_B,\phi_1)}{B(\phi_B)}
                               \Theta(\mu_\text{PS}^2-t)\mathcal{O}(\phi_R)\dif\phi_B\dif\phi_1\nonumber\\
  &+\int R(\phi_R)\mathcal{O}(\phi_R)\dif\phi_R. \label{eq:ps-cancellation}
\end{align}
To order $\alpha_s$ we have $\bar B= B$ and the first integral
cancels as required by the matching conditions, thus restoring the
fixed-order result. This explicitly demonstrates how variations in the
parton shower contributions induced by S-events are subtracted by the
MC@NLO subtraction terms $D$ in the definition of $H$.
Numerically, however, this cancellation can be severely spoiled, 
potentially leading to large deviations from the
fixed-order result. For the deviations to be significant the term on
the first line of equation \eqref{eq:ps-cancellation} must be similar in
size to the fixed-order term on the second line of
\eqref{eq:ps-cancellation}. One can therefore expect large deviations
from the fixed-order calculation only if \emph{both} of the following
conditions are met:
\begin{enumerate}
\item[1)] The factor $\bar B-B$ dressed with the parton shower
  splitting kernels ($\frac{D}{B}$ in \eqref{eq:ps-cancellation}) is
  comparable in magnitude relative to the real-emission matrix
  elements in $R$. This depends on the size of the NLO corrections
  that enter $\bar B$ and on the splitting kernels in the phase space
  region of hard emissions.
\item[2)] The phase space of interest is accessible to the parton
  shower so that the first integral in \eqref{eq:ps-cancellation} has
  support in that region. This depends on the choice of
  $\mu_\text{PS}$ and on the shower (through the functional form of
  $t(\phi_R)$).
\end{enumerate}

The formally sub-leading contributions originating from the parton
shower matching in the first integral in \eqref{eq:ps-cancellation}
are, to a large extent, controlled by the choice of $\mu_\text{PS}$.
To access the matching uncertainties we will therefore vary this
parameter by factors \num{2} and \num{0.5}. With two different parton
showers at our disposal we have an additional handle on these
uncertainties through the functional form of $t(\phi_R)$. The nominal
choice for $\mu_\text{PS}$ in the CS shower will be
$\mu_\text{PS}=m_{HH}/2$, in line with $\mu_R$ and $\mu_F$. As
outlined above, such a choice would open up the entire emission phase
space in case of the Dire shower. Our nominal choice for the Dire
shower will therefore be $\mu_\text{PS}=m_{HH}/4$, which allows us to
perform both up and downwards variations.

Based on the argument presented above, one might expect to see large
parton shower contributions in the high-$p_T$ tails of other processes with
large K-factors. However, it is important to note that for such effects to be
visible the $\bar{B}-B$ factor must remain large, relative to the real-emission 
matrix element, also when multiplied by the parton shower splitting kernels.
In single Higgs boson production through gluon fusion, for example, one
might anticipate a large shower contribution in the tail
of the Higgs transverse momentum spectrum due to the large NLO
K-factor. However, in this case the parton shower splitting kernels 
underestimate the real-emission matrix elements significantly, such
that the parton shower contributions in the tail are very small in an
MC@NLO-matched calculation \cite{Alioli:2008tz,Hoeche:2011fd}. For the
parton showers considered in this work, this holds even when taking
into account the full top mass dependence in the real-emission matrix
elements, which reduce the size of $R$ by more than an order of
magnitude in the tail of the transverse momentum distribution.

\section{Results}
\label{sec:results}

\subsection{Leading Order Results}
\label{sec:lops}

We start our discussion with predictions obtained in the most simple
setup, using leading order matrix elements for inclusive Higgs boson pair
production supplemented by a parton shower. This type of simulation
will be referred to as ``LO+PS'' in what follows. Since the transverse
momentum of the Higgs boson pair is zero at leading order, any non-zero value
of this observable is entirely generated by the parton shower. 
As a reference, we use a fixed-order
prediction obtained by simulating the process $p\ p\to H\ H\ j$
with leading order matrix elements.

\begin{figure}
\includegraphics[width=.5\linewidth]{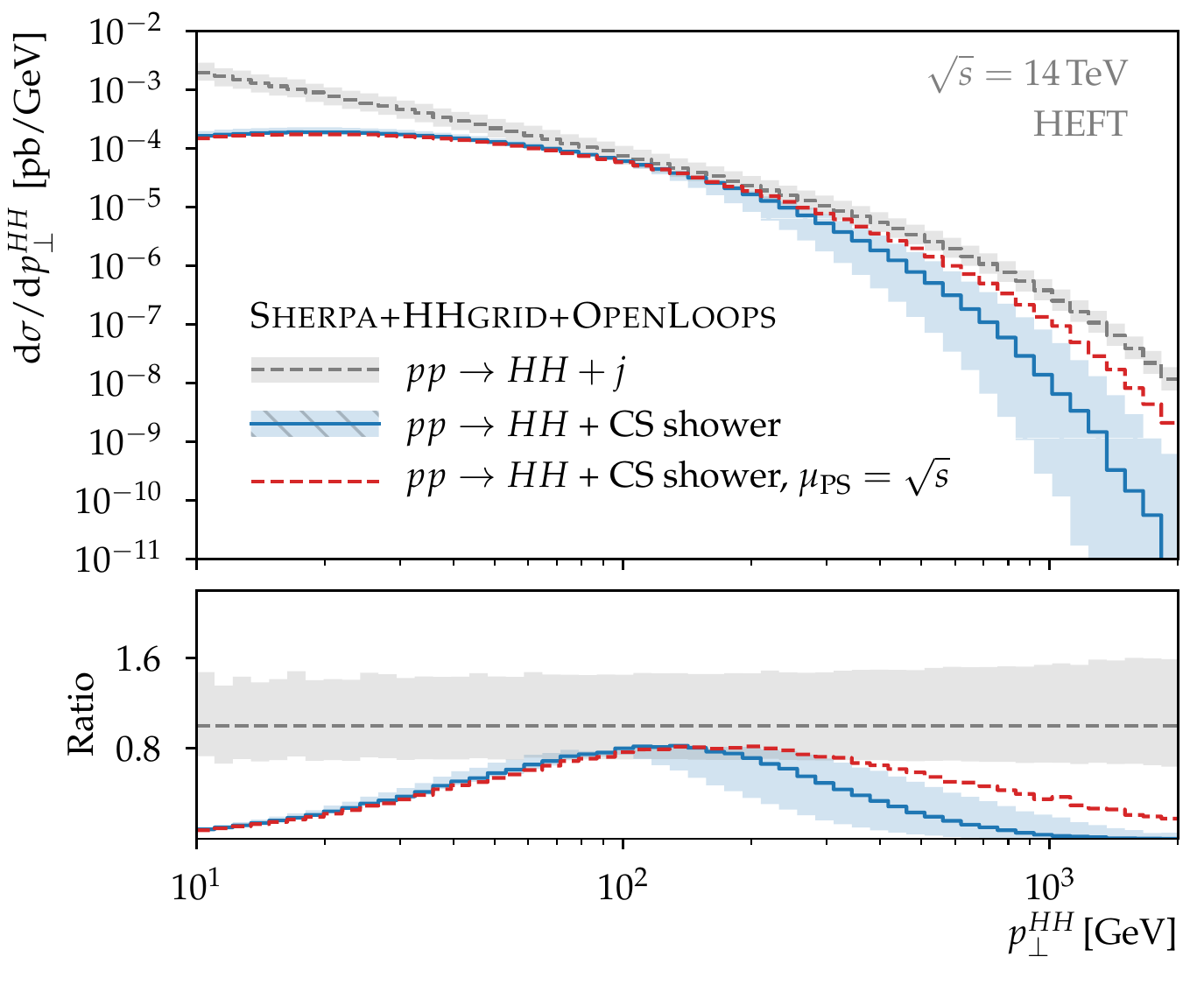}
\includegraphics[width=.5\linewidth]{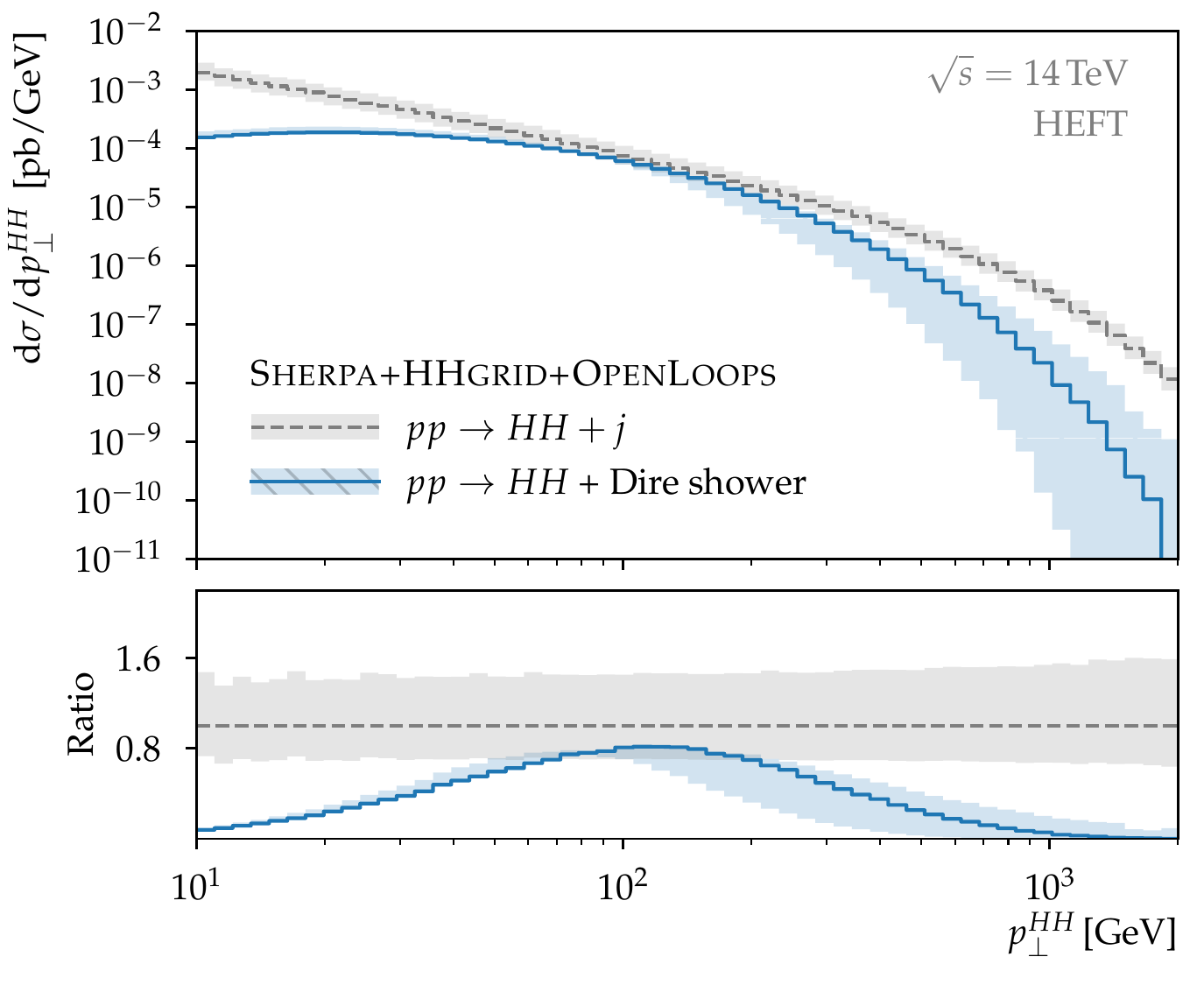}
\caption{Parton shower predictions for the $p_\perp^{HH}$ spectrum in
  a LO+PS type simulation compared to a fixed-order calculation in the
  HEFT approximation. The uncertainty band around the fixed-order
  result is obtained by varying $\mu_F$ and $\mu_R$.
  Uncertainties on the LO+PS results are obtained by varying
  $\mu_\text{PS}$.}
\label{fig:lops-heft}
\end{figure}

\begin{figure}
\includegraphics[width=.5\linewidth]{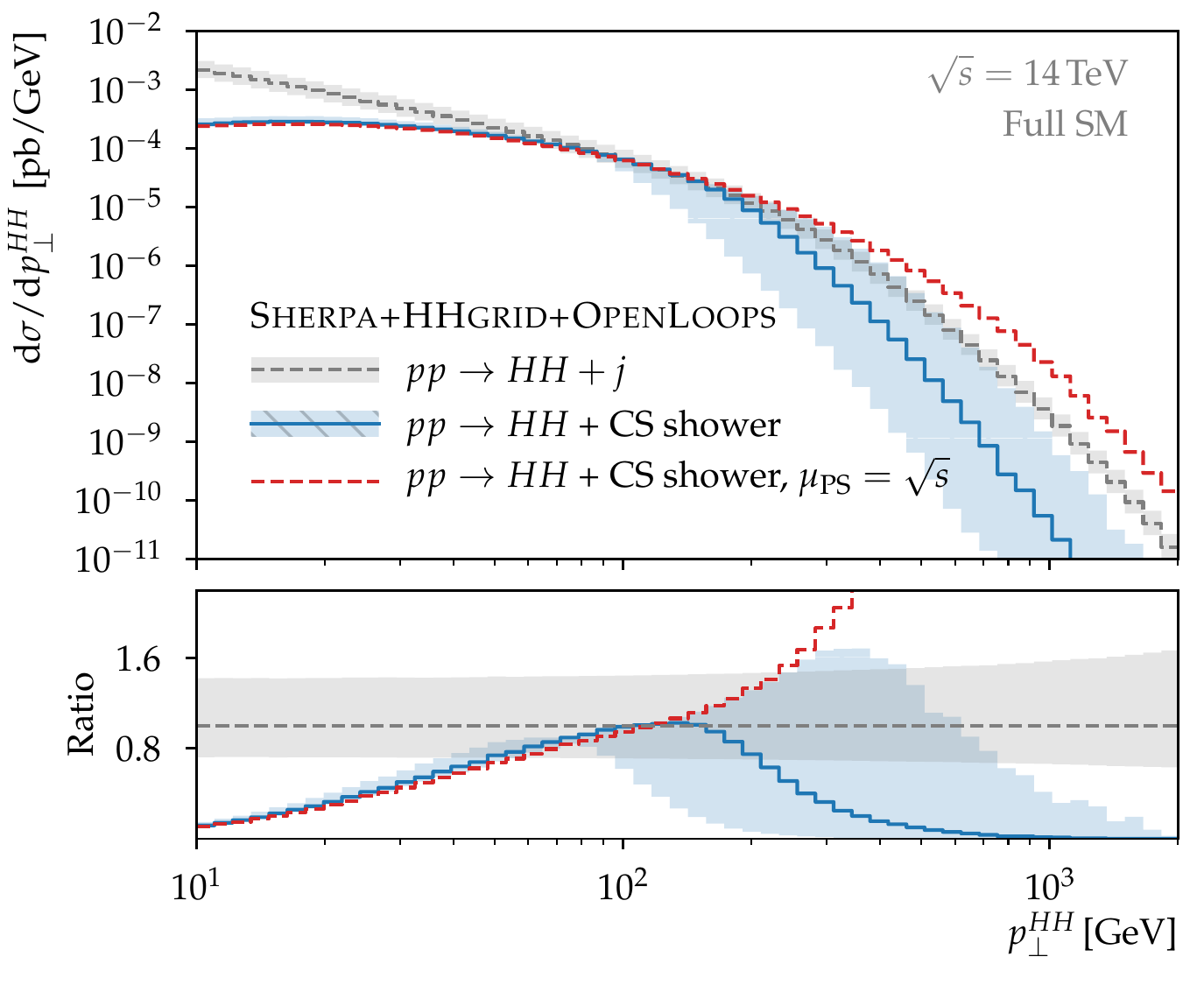}
\includegraphics[width=.5\linewidth]{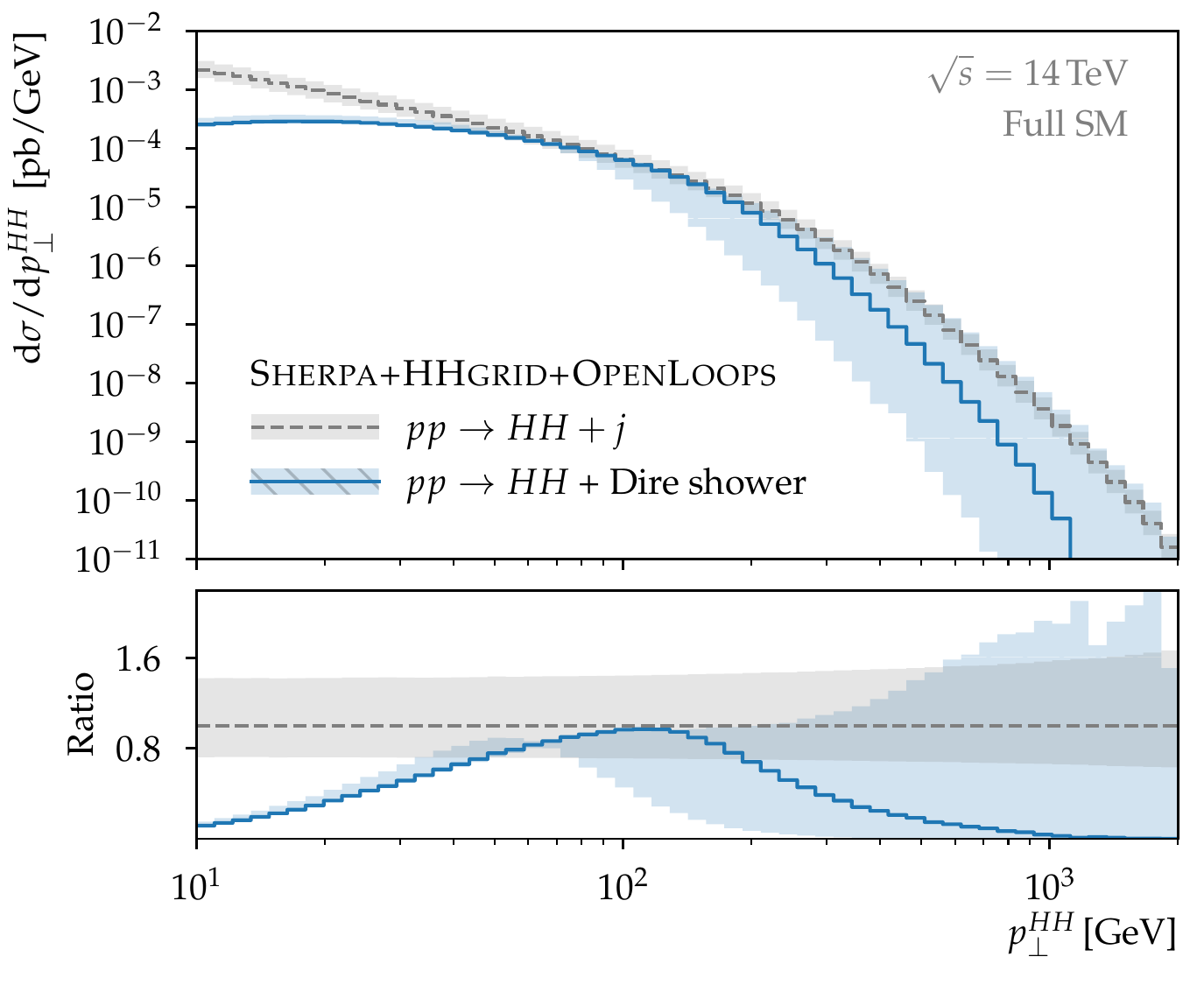}
\caption{Parton shower predictions for the $p_\perp^{HH}$ spectrum in
  a LO+PS type simulation compared to a fixed-order calculation in the
  full SM. The uncertainty band around the fixed-order result is
  obtained through variations of $\mu_F$ and $\mu_R$. Uncertainties on
  the LO+PS results are obtained by varying $\mu_\text{PS}$.}
\label{fig:lops-full}
\end{figure}

Figure~\ref{fig:lops-heft} and \ref{fig:lops-full} show the result of
our comparison both in the HEFT approximation and in the full theory,
respectively. Comparing the full SM and the HEFT approximation, we
observe qualitatively different parton shower effects. In the HEFT
approximation, both parton showers significantly underestimate the
fixed order spectrum in the tail of the distribution. Even if the full
phase space is made available to the showers they do not reproduce the
slowly falling transverse momentum spectrum predicted by the
fixed-order HEFT matrix elements. To show this for the CS shower we display
also LO+PS results obtained by setting the parton shower starting scale to
the hadronic center-of-mass energy $\mu_\text{PS}=\sqrt{s}$. In the case
of the Dire shower, the full phase space is already available for
$\mu_\text{PS}=m_{HH}/2$, which corresponds to the upper edge of the
uncertainty band.

In the full SM, by contrast, for large enough values of the parton shower
starting scale $\mu_\text{PS}$ both parton showers overestimate the 
fixed-order prediction. For the CS shower, this
effect is restricted to smaller transverse momenta, due to the choice
of evolution variable. If we lift any phase space restriction in the
CS shower, by setting $\mu_\text{PS}=\sqrt{s}$, we observe that
in the tail of the distribution the shower overestimates the fixed-order 
predictions by more than an order of magnitude. 
The upper edge of the Dire shower uncertainty band 
also overestimates the fixed-order prediction, although this feature  
is not as pronounced as for the CS shower.

It is therefore evident that the Born matrix elements dressed with
splitting kernels can strongly overestimate the real emission matrix
elements in the full SM and strongly underestimate the real emission
matrix elements in the HEFT. The former effect is, however, to a
certain extent limited by the phase space constraint implemented
through the parton shower starting scales.

Naively, the large differences between the HEFT and the full SM
simulations may seem surprising. However, the high-energy behaviour of
the HEFT real emission amplitudes is unphysical because the momentum
transfers vastly exceed the top-quark masses which have been
integrated out in the HEFT approximation. As a result, the spectrum
calculated at fixed-order falls off extremely slowly in the HEFT and
the parton shower kernels thus tend to underestimate the spectrum in
the tail. A similar slow fall off can been observed in the HEFT
approximation for the Higgs boson transverse momentum in single Higgs
boson
production~\cite{Baur199038,Ellis1988221,Buschmann:2014sia,Neumann:2016dny,Caola:2016upw}.

\subsection{NLO Results}

\begin{figure}
\includegraphics[width=.5\linewidth]{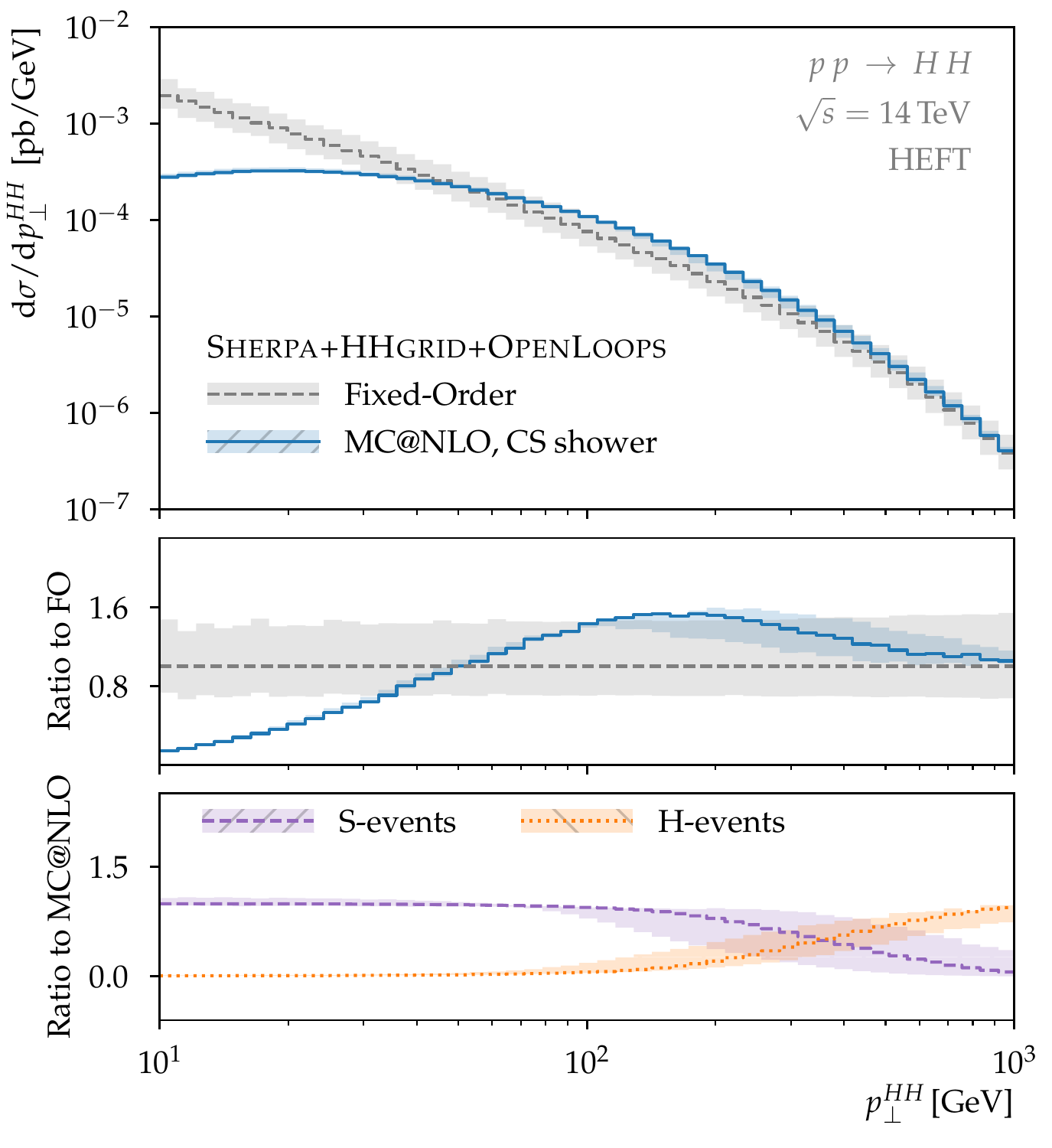}
\includegraphics[width=.5\linewidth]{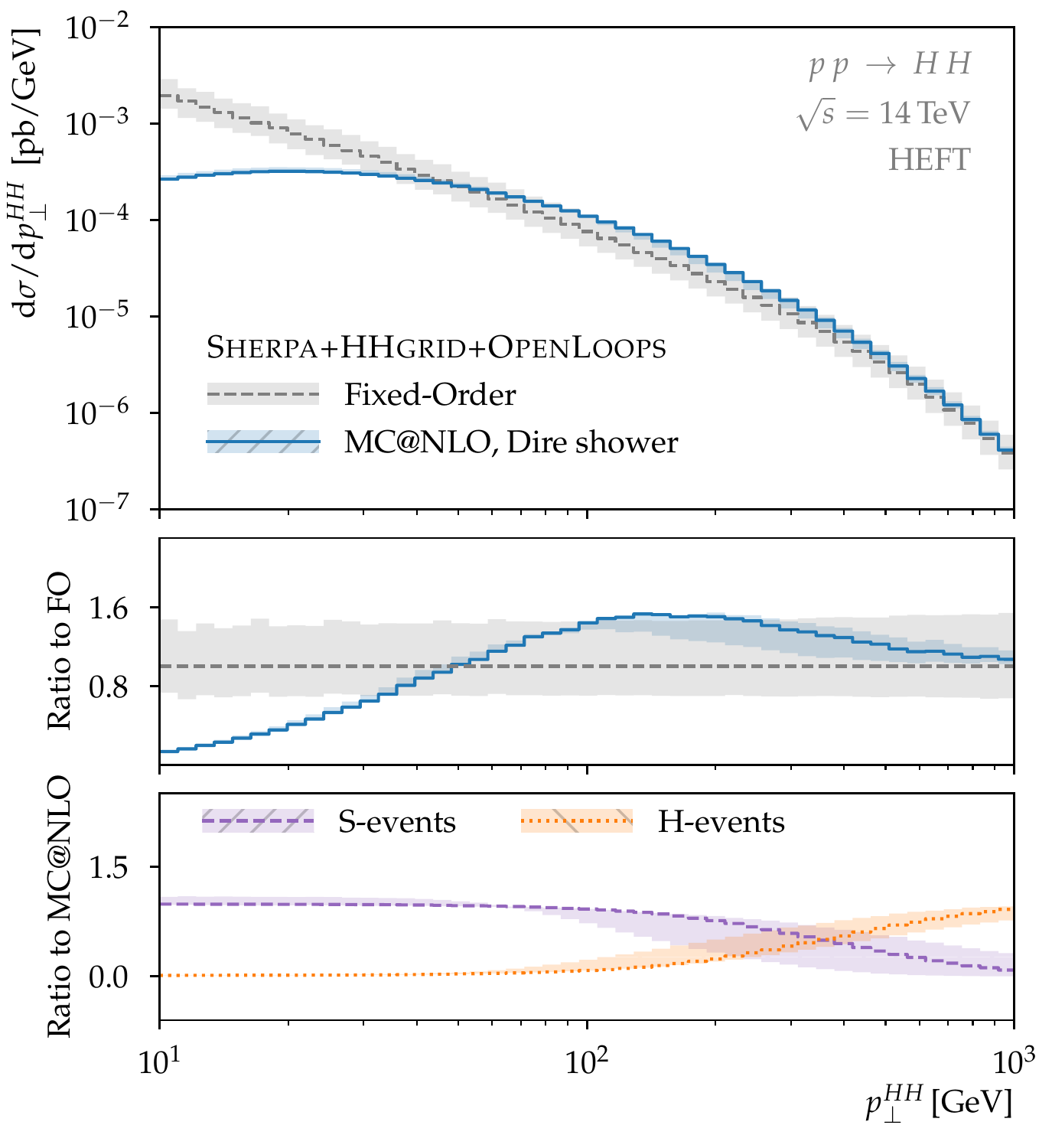}
\caption{Parton shower NLO matching effects on the $p_\perp^{HH}$
  spectrum in the HEFT approximation. The left panel shows results
  obtained with the CS shower, while the results on the right were
  generated with the Dire shower. The uncertainty band around the
  fixed-order result is obtained through variations of $\mu_F$ and
  $\mu_R$. Uncertainties on the MC@NLO predictions are obtained by varying
  $\mu_\text{PS}$.}
\label{fig:mcnlo-heft}
\end{figure}

We start the discussion of NLO-matched parton shower simulations with
the results of a HEFT treatment, shown in Figure~\ref{fig:mcnlo-heft}. As 
discussed in Section~\ref{sec:lops} (and shown in Figure~\ref{fig:lops-heft}) 
the combination of Born matrix elements and parton
shower splitting kernels strongly undershoot the full real-emission
matrix elements when employing the HEFT approximation. We therefore expect
the tail of the distributions to converge to the fixed-order result
both for the CS shower and the Dire shower. As shown in Figure~\ref{fig:mcnlo-heft}, 
this is indeed the case.
In addition to a more precise description of the tail (compared to
the LO+PS type simulations) we observe a reduction of the parton shower
starting scale uncertainties. The individual variations of H and
S-event contributions are of order one for some values of
$p_\perp^{HH}$ but cancel to a large extent in the sum.

\begin{figure}
\includegraphics[width=.5\linewidth]{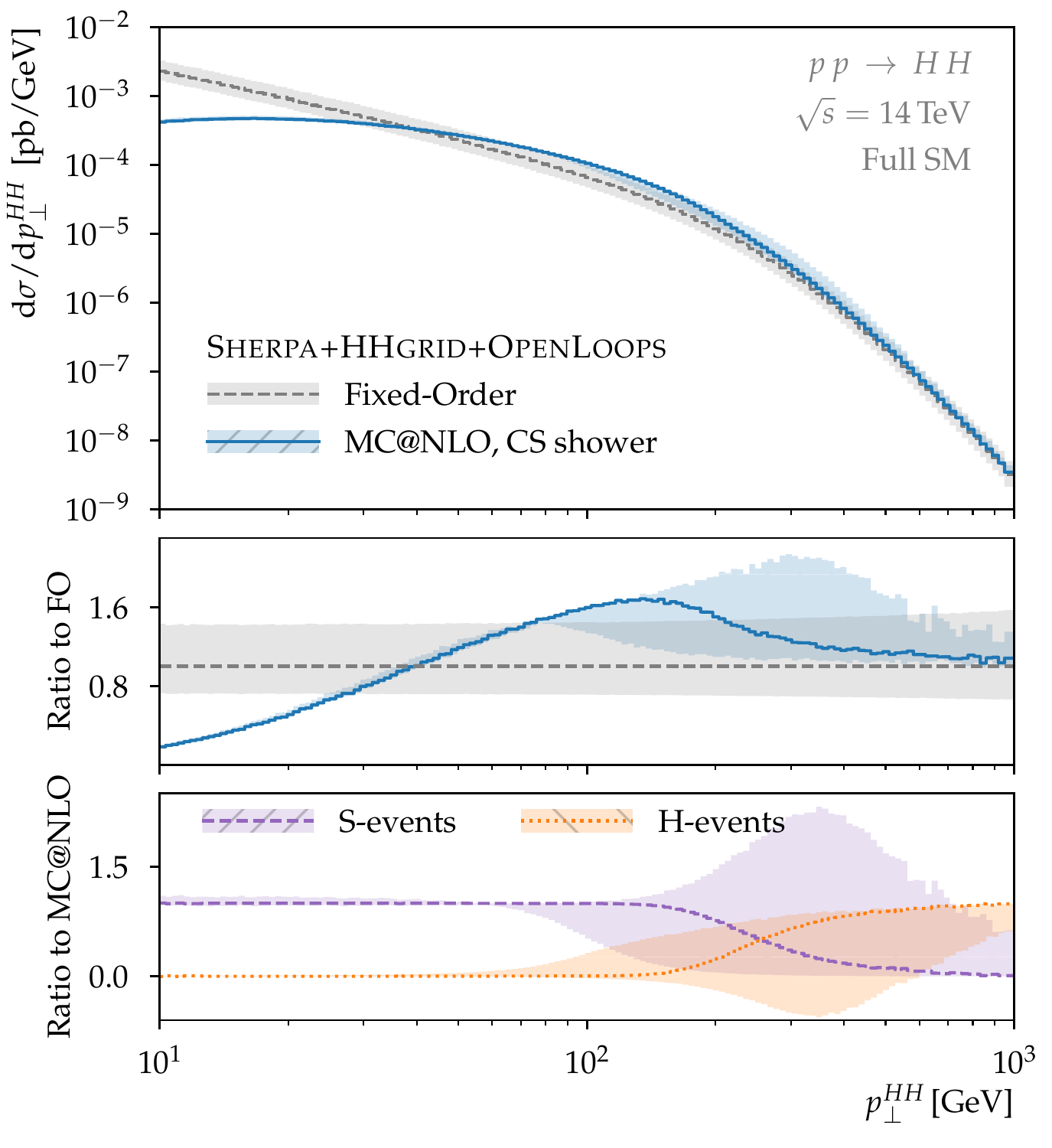}
\includegraphics[width=.5\linewidth]{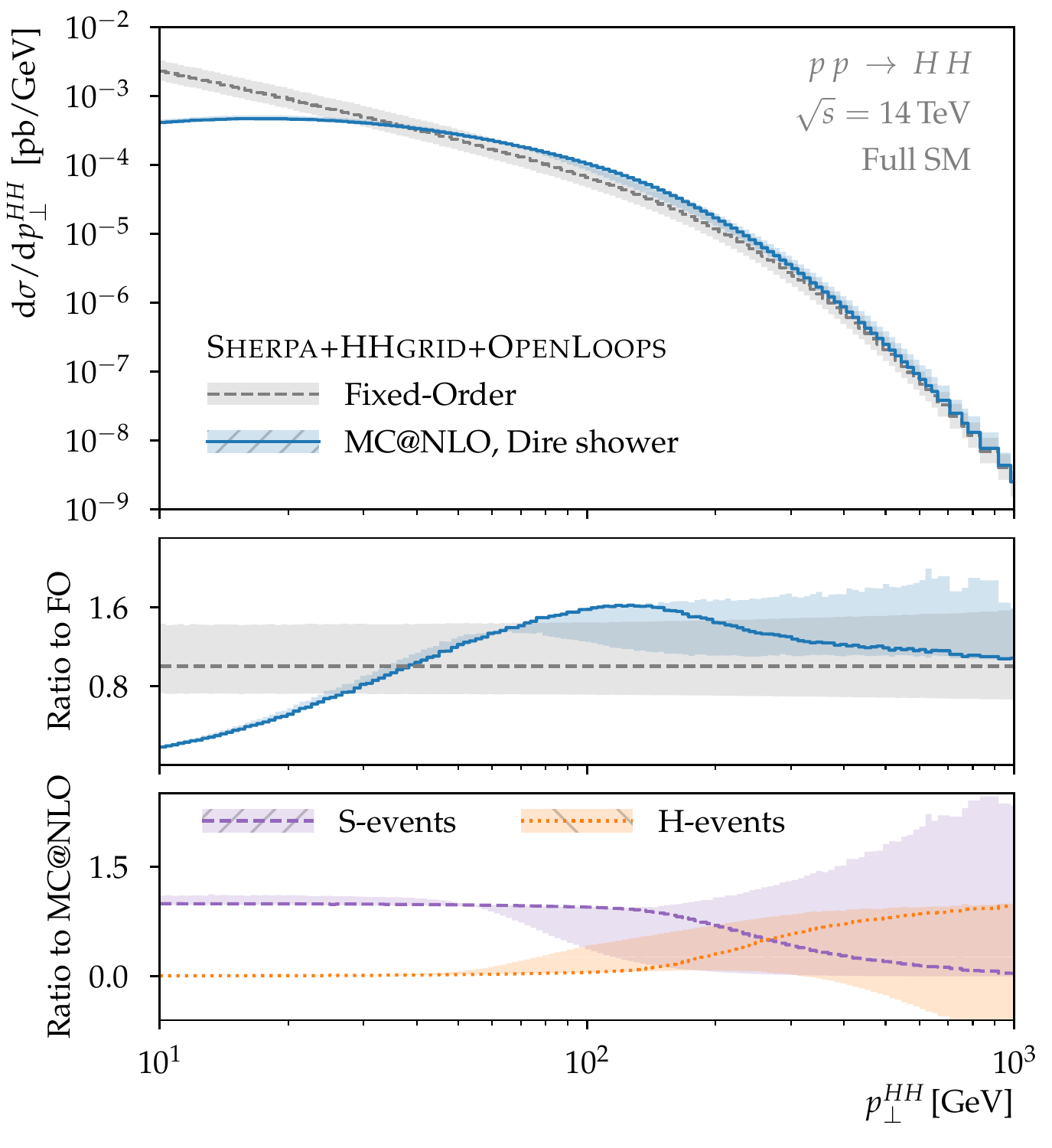}
\caption{Parton shower NLO matching effects on the $p_\perp^{HH}$
  spectrum in a full SM calculation. The left panel shows results
  obtained with the CS shower, while the results on the right were
  generated with the Dire shower. The uncertainty band around the
  fixed-order result is obtained through variations of $\mu_F$ and
  $\mu_R$. Uncertainties on the MC@NLO predictions are obtained by varying
  $\mu_\text{PS}$.}
\label{fig:mcnlo-full}
\end{figure}

Moving on to the discussion of results in the full SM, we remind the
reader of our findings in the corresponding LO+PS type simulations. In
the full SM, the parton shower splitting kernels in combination with
Born matrix elements overestimate the real-emission matrix elements
(see Figure~\ref{fig:lops-full}). The parton shower effects in the tail of the
$p_\perp^{HH}$ distribution are therefore large. As shown in Figure~\ref{fig:mcnlo-full}, 
this also holds at NLO. The parton shower
matched results converge to the fixed order result in the tail for
nominal choices for $\mu_\text{PS}$. Upward variation of
$\mu_\text{PS}$, however, (indicated by the upper edge of the blue
uncertainty bands) lead to parton shower effects of up to
+\SI{100}{\percent} even in the tail of the distribution. As shown in
the lower panels of Figure~\ref{fig:mcnlo-full}, the excesses in the
tail are indeed generated by parton shower emissions off S-events. The
extent of these effects is limited by the phase space available to the
parton shower, as determined by the choice of $\mu_\text{PS}$ and the
functional form of the evolution variable. We observe that results generated
using the Dire shower, particularly for larger values of $\mu_\text{PS}$,
have a different shape than those generated with the CS shower.

\begin{figure}
  \centering
  \includegraphics[width=.5\linewidth]{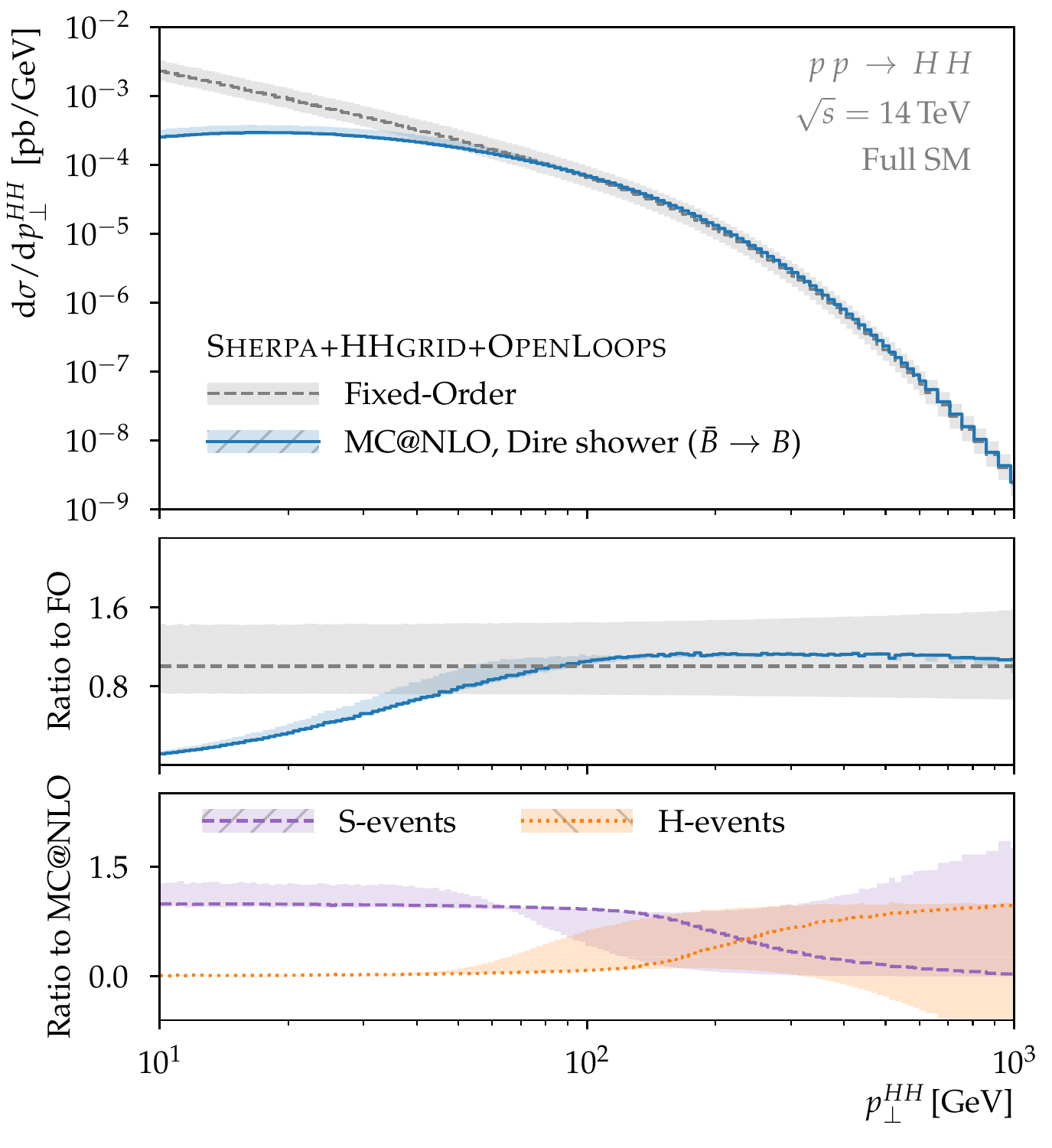}
  \caption{Parton shower effects on the $p_\perp^{HH}$ spectrum in a
    modified MC@NLO simulation where we replaced the S-event
    differential seed cross section $\bar B$ by $B$. The uncertainty
    band around the fixed-order result is obtained through variations
    of $\mu_F$ and $\mu_R$. Uncertainties on the MC@NLO predictions are
    obtained by varying $\mu_\text{PS}$.}
  \label{fig:mcnlo-dire-fix}
\end{figure}

For the MC@NLO algorithm, by construction, the large parton shower
effects in the tail should be cancelled to first order in $\alpha_S$.
As outlined in Section~\ref{sec:nlops_unc}, any mis-cancellation is due to
a numerically large discrepancy between $B$ and $\bar B$. We
demonstrate this explicitly in Figure~\ref{fig:mcnlo-dire-fix}, where
we show modified Dire MC@NLO with $B$ substituted for $\bar B$,
leading to a complete cancellation of the first integral in
\eqref{eq:ps-cancellation}. This procedure eliminates large parts of
the excess in the tail independently of $\mu_\text{PS}$, as anticipated.
Variations in S- and H-event contributions remain large, as shown in
the lower panels of Figure~\ref{fig:mcnlo-dire-fix}, but they cancel
in the sum. The procedure of replacing $\bar B$ with $B$ would, of
course, spoil the NLO accuracy of any inclusive observable but allows
us here to demonstrate the origin of the discrepancy between the 
showered and fixed-order results in the tail of the $p_\perp^{HH}$ distribution.

In the HEFT approximation one may naively expect effects of similar
size in the tail of the distributions. As demonstrated using a LO+PS
simulation and as shown in Figure~\ref{fig:lops-heft}, however, the
fixed-order real emission contributions completely dominate in this
region. The bulk of the contributions in the tail are hence generated
by the second integral of \eqref{eq:ps-cancellation}. As a result, the
relative impact of parton shower effects in the tail remains small as
shown in Figure~\ref{fig:mcnlo-heft}.

\subsection{Comparison to the Literature}

\begin{figure}
  \centering
  \includegraphics[width=.6\linewidth]{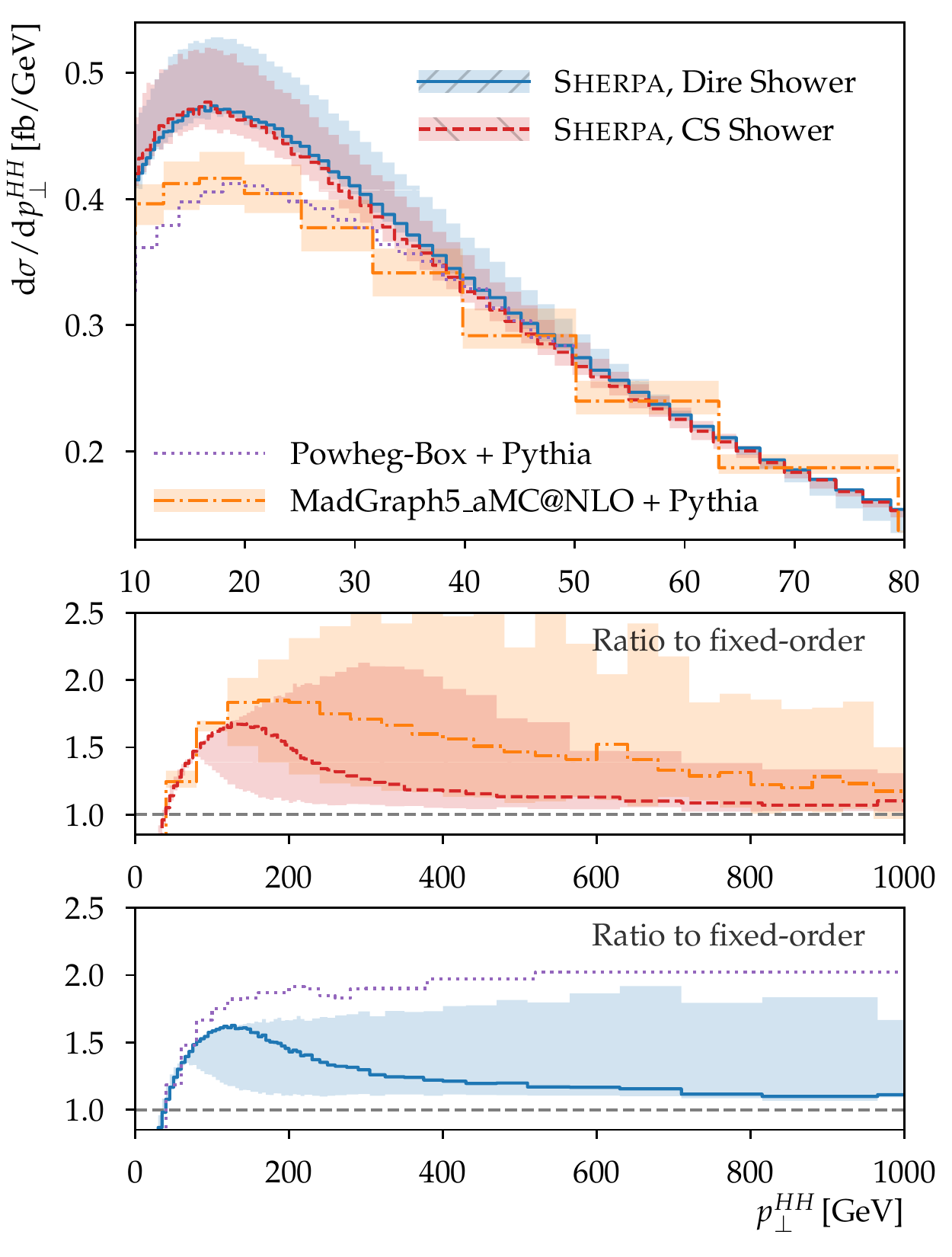}
  \caption{Comparison with NLO parton shower matched results from the
    literature. The lower panels show ratios to the fixed-order
    prediction and cover a wider range of $p_\perp^{HH}$ than the
    upper panel. The uncertainties on parton shower matched
    predictions were obtained by varying $\mu_\text{PS}$.}
  \label{fig:comp-lit}
\end{figure}

In Figure~\ref{fig:comp-lit} we compare our results for the
$p_\perp^{HH}$ spectrum to the NLO parton shower matched results
presented in reference \cite{Heinrich:2017kxx}. These results were
obtained with the Pythia 8 shower
\cite{Sjostrand:2007gs,Sjostrand:2014zea} interfaced to
MadGraph5\_aMC@NLO \cite{Alwall:2014hca,Hirschi:2015iia} and POWHEG
BOX \cite{Alioli:2010xd} for matching according to the MC@NLO method
and the POWHEG method \cite{Nason:2004rx}, respectively. In
MadGraph5\_aMC@NLO the nominal value of $\mu_\text{PS}$ is set
randomly in the interval $[0.1 H_T/2,H_T/2]$, where $H_T$ is the sum of
the transverse masses of the Higgs bosons.
For the simulations based on MadGraph5\_aMC@NLO
and Pythia we show uncertainty bands that were obtained by varying the
nominal parton shower starting scale by factors of \num{2} and
\num{0.5}. The POWHEG matching prescription can be recovered within
the MC@NLO framework by setting the parton shower starting scale to
the collider energy $\mu_\text{PS}=\sqrt{s}$ and by setting $D=R$
\cite{Hoeche:2011fd}. Therefore, lacking a natural equivalent to
$\mu_\text{PS}$ in the POWHEG framework, we compare only to nominal
POWHEG predictions produced with the \texttt{hdamp} parameter set 
to $\SI{250}{\giga\electronvolt}$, as described in \cite{Heinrich:2017kxx}.

Focusing on the region $p_\perp^{HH}>\SI{100}{\giga\electronvolt}$, we
note that all MC@NLO predictions considered here are generally
compatible within the uncertainty bands. However, the agreement between the
nominal results of our simulations and the fixed-order result is much
better in the tail. The POWHEG results exhibit a very large
excess in the tail that is not covered by the uncertainty bands of our
Sherpa predictions. Similar discrepancies between MC@NLO and POWHEG
have been observed in the context of other processes
\cite{Alioli:2008tz,Hoeche:2011fd} and can be attributed to the large
phase space available to the parton shower as a result of setting
$\mu_\text{PS}=\sqrt{s}$ and the numerically large discrepancy between
$\bar B$ and $B$ \cite{Nason:2012pr}. As described in Section~\ref{sec:ps}, 
the former can be achieved in Dire by setting
$\mu_\text{PS}=m_{HH}/2$, which is represented by the upper edge of
the uncertainty band around the Dire prediction. We note that the
shape of this curve is in fact most similar to the POWHEG prediction
in the tail of the distribution.

Comparing the different uncertainty bands themselves, we observe large
differences. The shape of uncertainty bands obtained with MadGraph5\_aMC@NLO
and with the CS shower are somewhat similar with a peak around
\SI{300}{\giga\electronvolt} but the size of the uncertainty band around the
MadGraph5\_aMC@NLO result is much larger throughout. The uncertainties
on the Dire prediction describe a more evenly shaped band.

Differences in the region of small transverse momenta are not fully
covered by the $\mu_\text{PS}$-variation bands. Although we expect
these variations to be indicative of NLO-matching uncertainties, we do
not expect them to cover all parton shower uncertainties. These
include, but are not limited to, ambiguities in the choice of a
kinematic recoil scheme and ambiguities in the choice of the
renormalization scales for the strong coupling in the splitting
kernels.

\begin{figure}
  \centering
  \includegraphics[width=.6\linewidth]{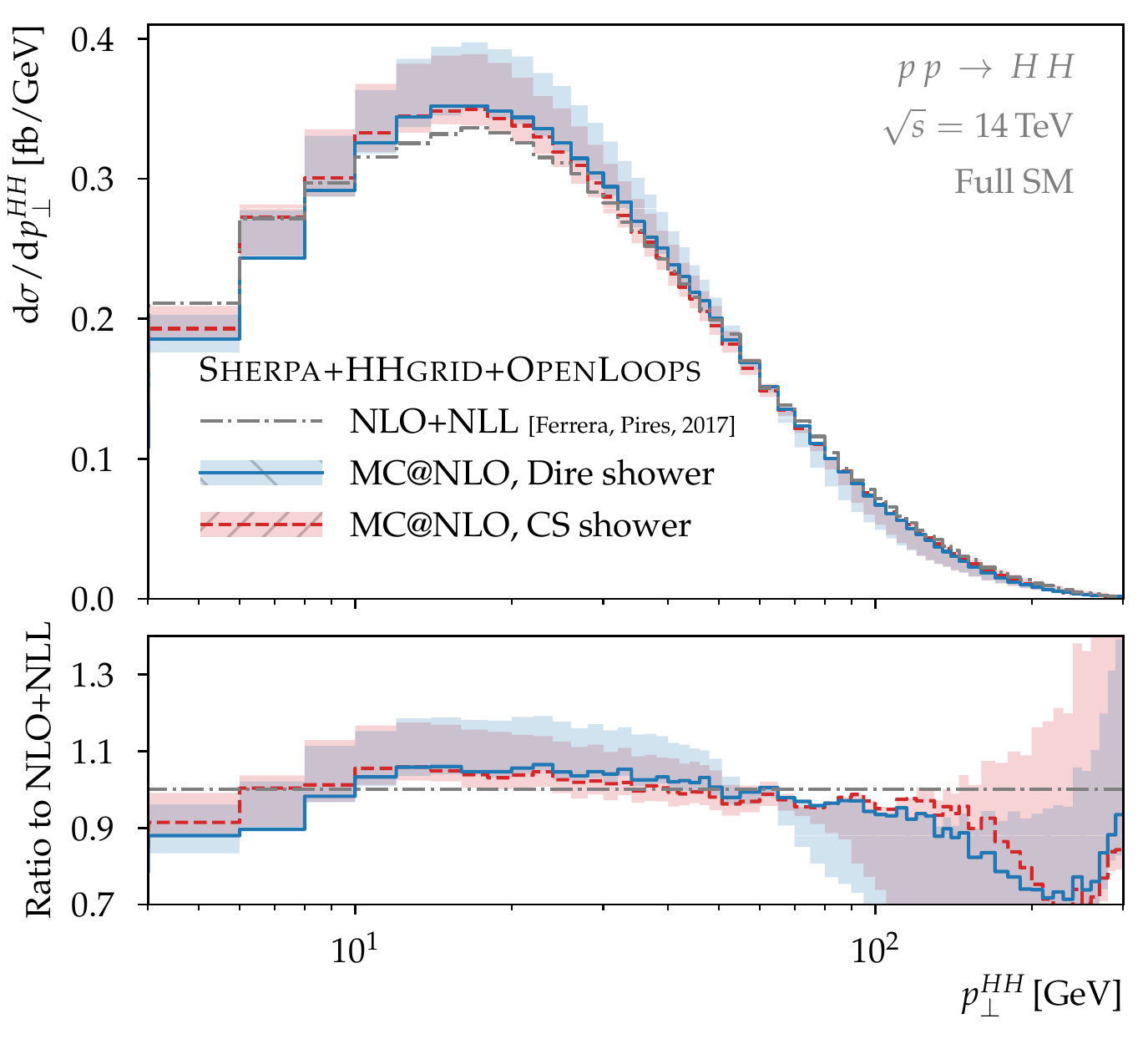}
  \caption{Comparison with results of the analytic resummation
    calculation of reference \cite{Ferrera:2016prr}. Uncertainties on
    the NLO parton shower matched results were obtained by varying
    $\mu_\text{PS}$.}
  \label{fig:comp-lit-ana}
\end{figure}

In Figure~\ref{fig:comp-lit-ana} we show a comparison to the
calculation of reference \cite{Ferrera:2016prr} which employed
analytic next-to-leading-log (NLL) resummation techniques instead of
parton showers. We observe good agreement within the uncertainties
except near the peak region and the region around
$p_\perp^{HH}=\SI{100}{\giga\electronvolt}$ where we find
discrepancies of about \SI{5}{\percent} that are not fully covered by
our uncertainty bands. Taking into account the resummation scale
uncertainties on the analytic results (not shown in Figure~\ref{fig:comp-lit-ana}), 
which are of the order of \SI{3}{\percent} in
the peak region and \SI{10}{\percent} near
$p_\perp^{HH}=\SI{100}{\giga\electronvolt}$ \cite{Ferrera:2016prr}, we
consider the observed agreement satisfactory.

\subsection{Other Observables}

Having discussed the Higgs boson pair transverse momentum at length, we
briefly discuss parton shower effects on a number of other
observables.

Lorentz invariant observables that depend only on the momenta of the
Higgs bosons are not affected by the parton shower. The kinematics of the
Higgs boson pair is only altered by emissions off dipoles spanned by two
initial state partons. The recoil generated by such an emission
affects the Higgs boson pair only through a Lorentz boost.
Lorentz-invariant quantities like the Higgs boson pair invariant mass are
therefore not affected by the parton shower. Our
simulations were checked by inspecting the Higgs boson pair invariant mass distributions,
we observe agreement within the statistical uncertainties of well
below one percent.

\begin{figure}
  \centering
  \includegraphics[width=.48\linewidth]{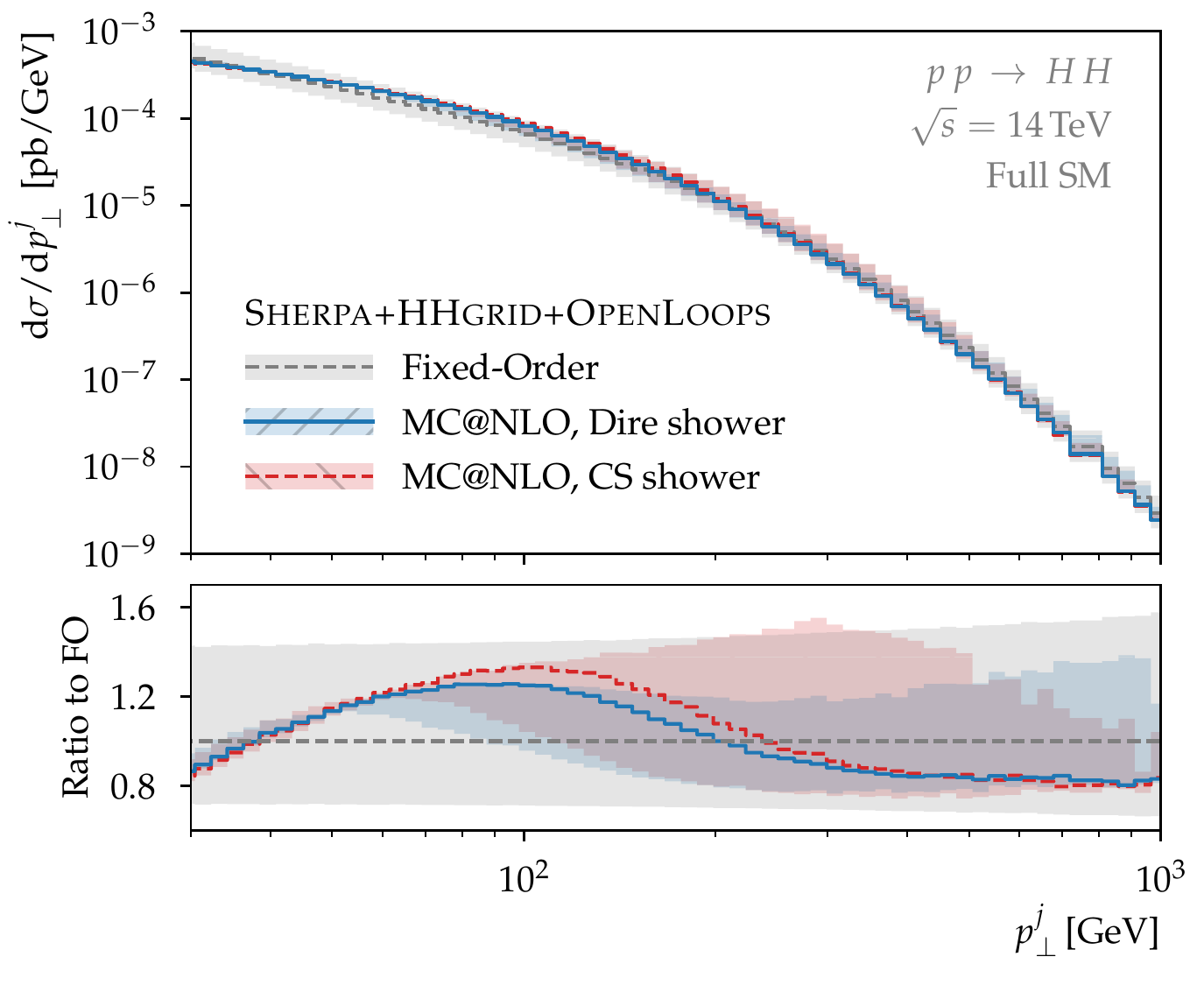}
  \includegraphics[width=.48\linewidth]{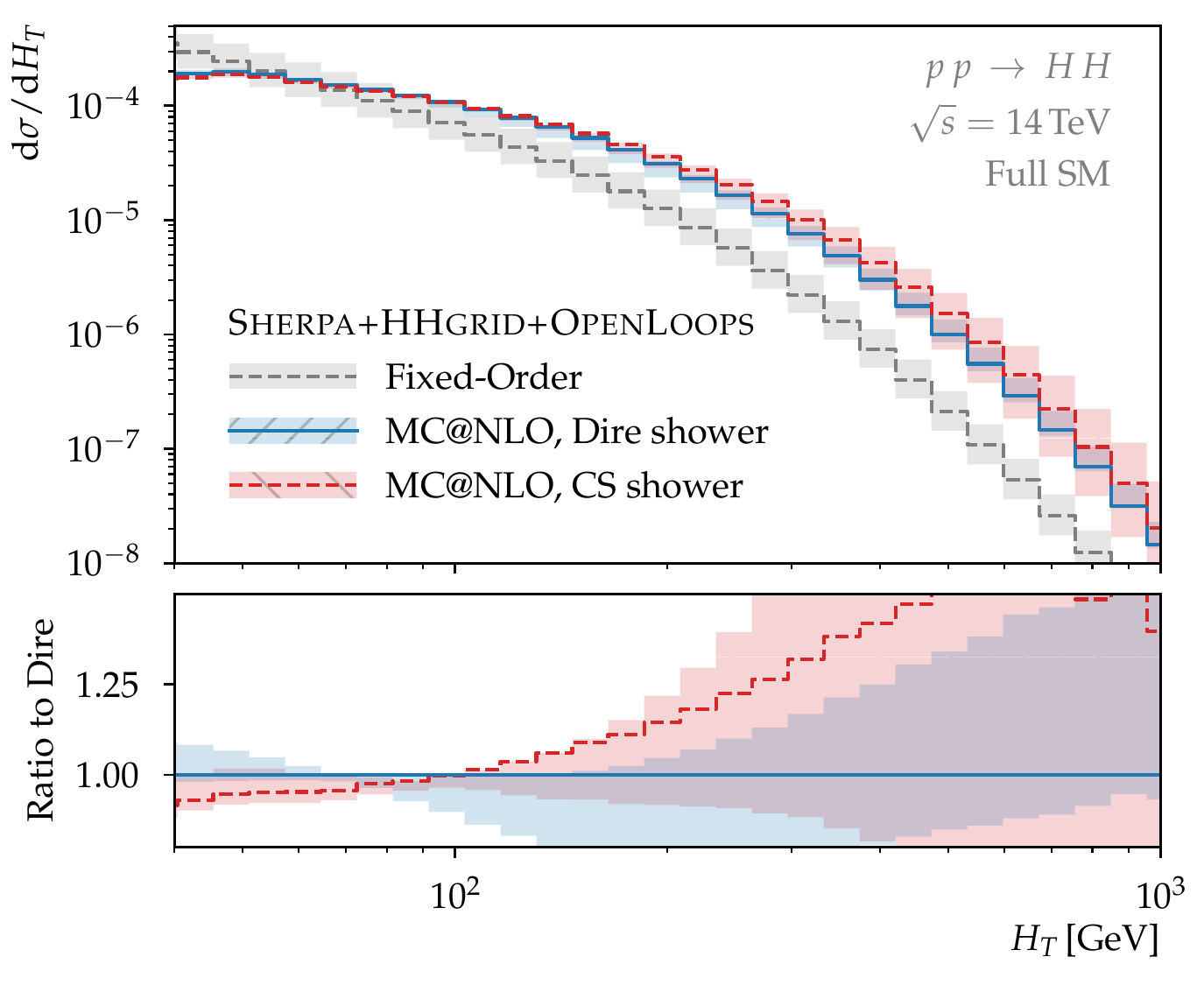}
  \caption{Parton shower effects on the leading jet transverse
    momentum (left panel) and on the scalar sum of jet transverse
    momenta. The uncertainty band around the fixed-order result is
    obtained through variations of $\mu_F$ and $\mu_R$. Uncertainties
    on the MC@NLO results are generated by varying $\mu_\text{PS}$.}
\label{fig:j-pt}
\end{figure}

Figure~\ref{fig:j-pt} shows the leading jet transverse momentum
$p_\perp^j$. As opposed to the Higgs boson pair, the parton emitted in the
hardest emission off an S-event and the final state parton in an H
event is affected strongly by secondary emissions because the  
kinematics are altered by final-final and final-initial dipoles. Such
emissions decorrelate the leading jet and the Higgs boson pair momenta.
Parton shower effects are therefore qualitatively different. 
The effect of the parton shower are generally moderate and 
the spectrum remains compatible with the fixed-order prediction within
the scale uncertainties. It is worth
noting that even the full $\mu_\text{PS}$-variation bands remain
within the uncertainty bands of the fixed-order calculation.

This picture drastically changes when considering observables that are
more sensitive to high multiplicity final states. As an example we
consider the differential $H_T$ distribution, defined as the scalar
sum of jet transverse momenta:
\begin{align*}
 H_T = \sum_i p_\perp^{j_i}\ ,
\end{align*}
where the index $i$ labels all jets in the respective event. In a
parton shower event, the total energy is typically distributed among a
larger number of jets than in a fixed-order calculation. Their
scalar contributions to $H_T$ are added, giving
rise to larger values of $H_T$ than for the fixed-order NLO calculation.
We show this effect in Figure~\ref{fig:j-pt}. Comparing the Dire and
CS shower predictions, we note that the uncertainty bands overlap but
that the shower starting scale variations are much larger for the CS
shower.

\begin{figure}
  \centering
  \includegraphics[width=.48\linewidth]{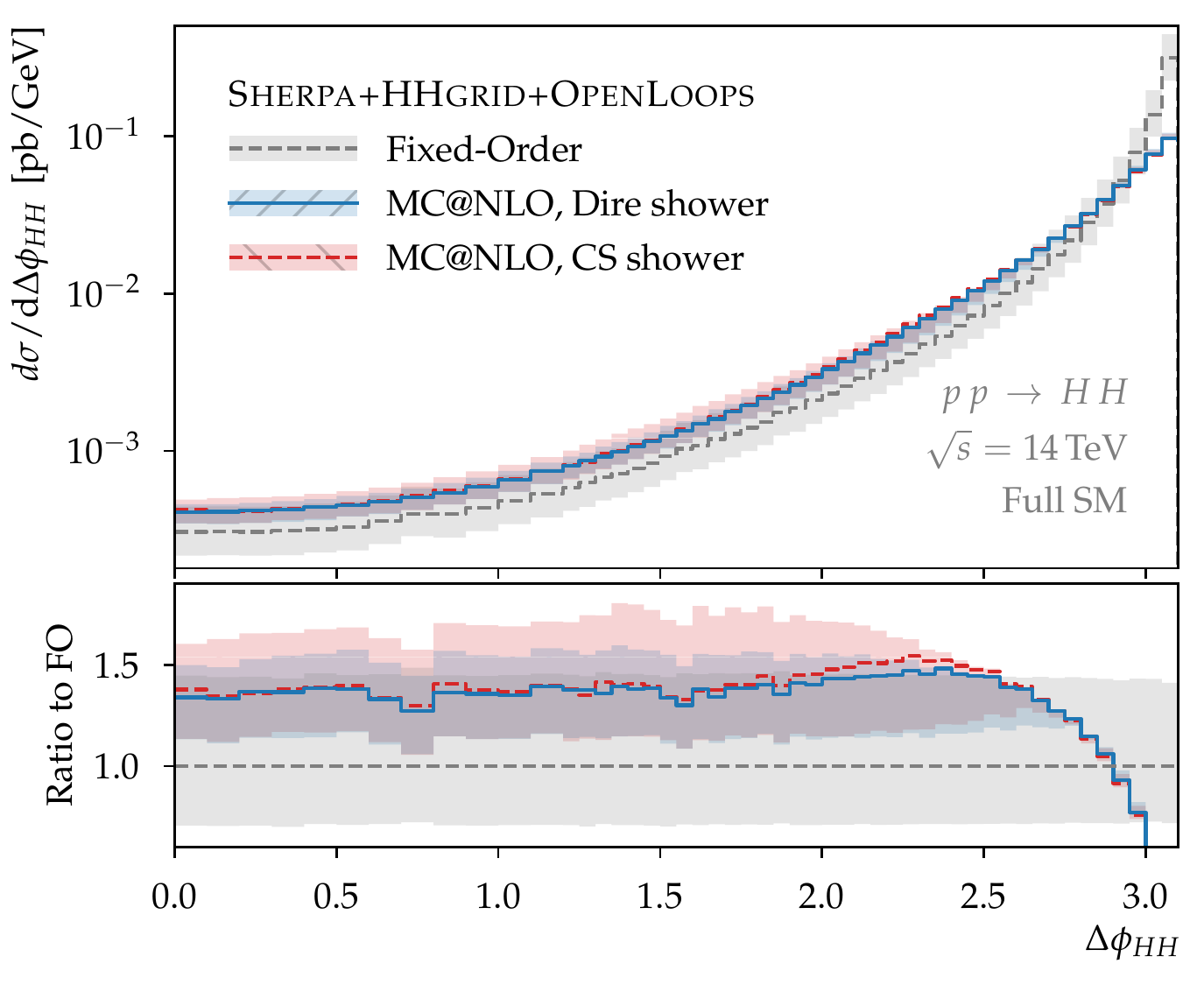}
  \includegraphics[width=.48\linewidth]{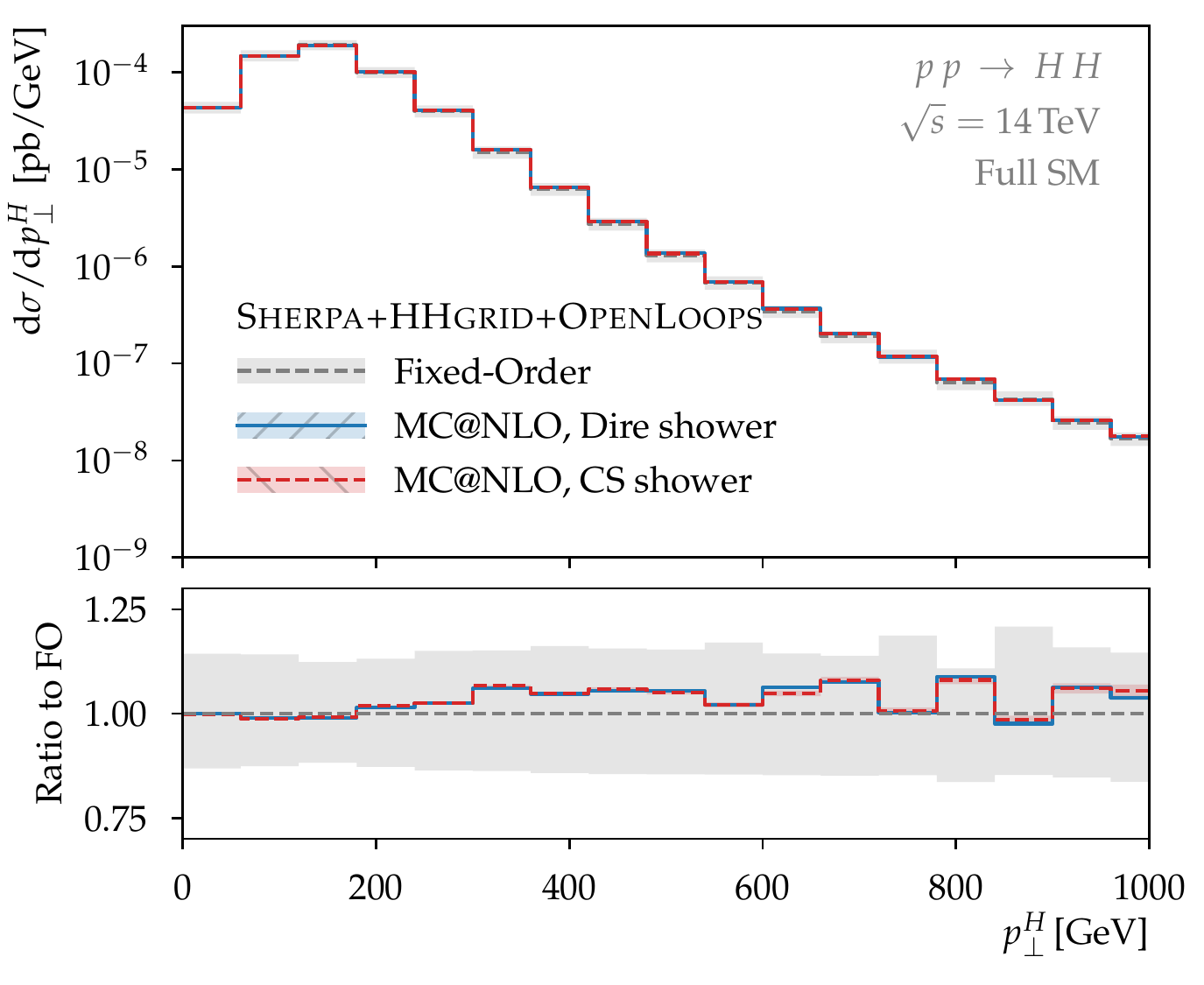}
  \caption{Parton shower effects on the azimuthal separation between
    the Higgs bosons (left panel) and on the transverse momentum of a
    randomly selected Higgs boson (right panel). The uncertainty band
    around the fixed-order result is obtained through variations of
    $\mu_F$ and $\mu_R$. Uncertainties on the MC@NLO predictions are
    generated by varying $\mu_\text{PS}$.}
\label{fig:h-obs}
\end{figure}

In Figure~\ref{fig:h-obs} we show the azimuthal separation between the Higgs bosons 
$\Delta\phi_{HH}$. At leading order, the momenta of the
Higgs bosons are perfectly correlated due to momentum conservation.
Only in events with additional radiation can one observe a non-trivial
distribution of the azimuthal separation between the Higgs bosons. As shown
in Figure~\ref{fig:h-obs}, parton shower corrections to the
fixed-order result are mostly covered by the fixed-order uncertainties
except in the region of $\Delta\phi_{HH}=\pi$ which corresponds to
back-to-back configurations and which is sensitive to soft QCD
emissions.

Also shown in Figure~\ref{fig:h-obs} is the transverse momentum of a
randomly chosen Higgs boson. The effect of a parton shower emission on
the transverse momentum of a given Higgs boson is random, either decreasing
or increasing its value. If the distribution was completely flat, any
parton shower effects would therefore average out. Since the
distribution is falling, the parton shower effects of
increasing the transverse momenta of low-$p_\perp$ Higgs bosons is not
counter-balanced by the effect of decreasing the transverse momenta of
high-$p_\perp$ Higgs bosons, thus inducing a slope relative to the
fixed-order result. This effect is small but clearly visible in Figure~\ref{fig:h-obs}.

\section{Conclusions}
\label{sec:conclusions}

We have presented a study of NLO parton shower matching uncertainties in
Higgs boson pair production through gluon fusion at the LHC. We
assessed these uncertainties by matching the fixed-order NLO
calculation to two dipole shower algorithms in the Sherpa event
generator according to the MC@NLO framework. The interplay between
fixed-order real emission contributions and parton shower emissions
was studied in detail through variations of the parton shower starting
scale. We find large matching uncertainties that exceed the
fixed-order uncertainties even in regions of phase space where the
fixed-order calculation is well motivated and where parton shower
matching effects are expected to be small. Our nominal predictions are
in good agreement with the fixed-order result in these regions,
however. A comparison to MC@NLO matched results from the literature
revealed qualitative differences which are, nevertheless, compatible
within the large uncertainties. We observe larger differences in a
comparison to POWHEG predictions in the tail of the transverse
momentum spectrum, where POWHEG overestimates the fixed-order spectrum
by a factor of \num{2}. We find reasonable agreement throughout
between our results and those obtained through analytic resummation
techniques.

\section*{Acknowledgements}

We would like to thank Stefan H{\"o}che for providing analytic results
for the subtraction and matching terms required in the interface to
the Dire shower, for technical assistance in the implementation, and
for helpful discussions. We thank Eleni Vryonidou for evaluating
the MadGraph5\_aMC@NLO + Pythia parton shower uncertainties and 
Matthias Kerner for his advice
regarding the use of the fixed-order NLO results. We would also like
to thank Jonas Lindert and Philipp Maierh{\"o}fer for their support
with OpenLoops. We are indebted to Gudrun Heinrich, Joey Huston,
Matthias Kerner, and Frank Krauss for their valuable comments on the
draft. This work is supported in part by the U.S. Department of Energy
under contract number DE-AC02-76SF00515 and by the Research Executive
Agency (REA) of the European Union under the Grant Agreement
PITN-GA2012316704 (HiggsTools).

\appendix

\cleardoublepage



\cleardoublepage

\bibliography{./bibliography}
\bibliographystyle{plain}

\end{document}